\documentclass[aps,prl,eprint,superscriptaddress,nofootinbib,notitlepage,showpacs]{revtex4-1}

\usepackage{epsfig,verbatim}
\usepackage{subfigure}
\usepackage{amsmath, amssymb, graphics}
\usepackage{color}
\usepackage{slashed}            % for slashed characters in math mode
\usepackage{bbm}                % for \mathbbm{1} (unit matrix)
\usepackage{units}              % for \unit[AMOUNT]{UNIT}
\usepackage{xspace}             % for 'smart' spaces after mathematics
\usepackage{bm}

\newcommand{\mathsym}[1]{{}}

\newcommand{\mode}[1]{\ensuremath{\mathcal{#1}}\xspace}

\newcommand\eps{\epsilon}

\newcommand{\codename}[1]{\textsc{#1}\xspace}
\newcommand{\class}{\codename{CLASS}}
\newcommand{\montepython}{\codename{Monte Python}}

\newcommand{\lcdm}{$\Lambda$CDM\xspace}

\begin{document}
%\pagestyle{headings}

%\preprint{}

\title{Localized correlated features in the CMB power spectrum and primordial bispectrum from a transient reduction in the speed of sound}
% PRL title
%\title{Transient reductions of the inflaton speed of sound in the Planck data}

\author{Ana Ach\'ucarro}
\email[]{achucar@lorentz.leidenuniv.nl}
\affiliation{Instituut-Lorentz for Theoretical Physics, Universiteit Leiden, 2333 CA Leiden, The Netherlands}
\affiliation{Department of Theoretical Physics, University of the Basque Country, 48080 Bilbao, Spain}
\author{Vicente Atal}
\email[]{atal@lorentz.leidenuniv.nl}
\affiliation{Instituut-Lorentz for Theoretical Physics, Universiteit Leiden, 2333 CA Leiden, The Netherlands}
\author{Pablo Ortiz}
\email[]{ortiz@lorentz.leidenuniv.nl}
\affiliation{Instituut-Lorentz for Theoretical Physics, Universiteit Leiden, 2333 CA Leiden, The Netherlands}
\affiliation{Nikhef, Science Park 105, 1098 XG Amsterdam, The Netherlands}
\author{Jes\'us Torrado}
\email[]{torradocacho@lorentz.leidenuniv.nl}
\affiliation{Instituut-Lorentz for Theoretical Physics, Universiteit Leiden, 2333 CA Leiden, The Netherlands}



%\thanks{}


%\date{\today}

%\renewcommand{\abstractname}{}
\begin{abstract}
% We perform a search for localized oscillatory features in the Planck CMB power spectrum, assuming they are caused by a transient reduction in the speed of sound of the adiabatic mode during effectively single-field, uninterrupted slow-roll inflation. We find several fits, for which we calculate the expected correlated signal in the primordial bispectrum, and compare it to the search for scale dependent bispectrum features carried out by the Planck collaboration.
% Where both searches overlap, we reproduce the Planck results reasonably well.
% In addition, some of our best fits lie outside the scales and frequency ranges searched by Planck, which calls for an extension in frequencies and envelopes of the templates used in Planck's search.
% By exploiting correlations between different observables, our results strongly suggest that current data might already be sensitive enough to detect transient reductions in the speed of sound as mild as a few percent, opening a new window for the presence of extra degrees of freedom during inflation.
% \\\\
The first year of observations by the Planck satellite mission shows that the cosmic microwave background (CMB) fluctuations are consistent with gaussian statistics in the primordial perturbations, a key prediction of the simplest models of inflation. However, there are hints of anomalies in the CMB power spectrum and bispectrum. We check for the possibility that some of these anomalous features have a common physical origin in a transient reduction of the inflaton speed of sound. We do this by exploiting predicted correlations between the power spectrum and bispectrum. Our results suggest that current data might already be sensitive enough to detect transient reductions in the speed of sound as mild as a few percent. Since this is a signature of interactions, it opens a new window for the detection of extra degrees of freedom during inflation.
\end{abstract}

%\keywords{}


\maketitle
%\thispagestyle{empty}


%%%%%%%%%%%%%%%%%%%%%%%%%%%%%%%%%%
%%%%%%%%%%%%%%%%%%%%%%%%%%%%%%%%
%%%%%%%%%%%%%%%%%%%%%%%%%%%%%%%%%%%

%\tableofcontents

%%%%%%%%%%%%%%%%%%%%%%%%%%%%%%%%%%
%%%%%%%%%%%%%%%%%%%%%%%%%%%%%%%%
%%%%%%%%%%%%%%%%%%%%%%%%%%%%%%%%%%%

The paradigm of inflation \cite{Guth:1980zm,Starobinsky:1979ty,Starobinsky:1982ee,Sato:1980yn,Linde:1981mu,Albrecht:1982wi} in its simplest realizations is consistent with the latest data releases from the Planck \cite{Ade:2013ktc} and WMAP \cite{Bennett:2012zja} satellites.
However, hints of a primordial oscillatory signal in the CMB bispectrum \cite{Ade:2013ydc} and of anomalies in the CMB power spectrum \cite{Bennett:2012zja,Ade:2013uln} motivate a search for correlated features produced by inflationary scenarios beyond canonical single-field.\footnote{By canonical single-field we mean slow-roll regime, Bunch-Davies vacuum and canonical kinetic terms.} Such correlation is in general expected and will differ depending on its physical origin \cite{Cheung:2007st}, so it can be used to discriminate among inflationary mechanisms.

On the theory side, several mechanisms that produce oscillatory features are being investigated. As first noted in \cite{Starobinsky:1992ts}, a step in the inflaton potential causes features in the spectra \cite{Wang:1999vf,Adams:2001vc,Gong:2005jr,Ashoorioon:2006wc,Chen:2008wn,Arroja:2011yu,Martin:2011sn,Adshead:2011jq,Arroja:2012ae,Takamizu:2010xy,Bartolo:2013exa}, and novel methodologies have been developed in \cite{Stewart:2001cd,Choe:2004zg,Dvorkin:2009ne,Adshead:2011bw,Miranda:2012rm,Adshead:2013zfa} for more generic transient slow-roll violations. The effect of a variable speed of sound has also been analyzed both in the power spectrum \cite{Achucarro:2010da,Hu:2011vr,Achucarro:2012fd} (for sudden variations see \cite{Park:2012rh,Miranda:2012rm,Nakashima:2010sa,Bean:2008na,Bartolo:2013exa}) and bispectrum \cite{Achucarro:2012fd,Adshead:2013zfa,Ribeiro:2012ar} (see \cite{Bean:2008na,Bartolo:2013exa} for sudden variations). Different initial vacuum states (see e.g. \cite{Danielsson:2002kx,Greene:2004np,Meerburg:2009ys,Jackson:2010cw}) or multi-field dynamics \cite{Gao:2012uq,Gao:2013ota,Saito:2013aqa,Noumi:2013cfa} may also cause oscillations in the primordial spectra.

On the observational side, searches in the CMB power spectrum data have been performed for a variety of scenarios, such as transient slow-roll violations \cite{Covi:2006ci,Adshead:2011jq,Benetti:2011rp,Adshead:2012xz,Benetti:2013cja,Hamann:2007pa,Benetti:2012wu,Miranda:2012rm}, superimposed oscillations in the primordial power spectrum \cite{Martin:2003sg,Flauger:2009ab,Aich:2011qv,Meerburg:2011gd,Peiris:2013opa,Meerburg:2013cla,Meerburg:2013dla} and more general parametric forms (see \cite{Ade:2013uln} and references therein). In addition, the Planck collaboration searched for features in the CMB bispectrum for a number of theoretically motivated templates \cite{Ade:2013ydc}. In none of these cases the statistical significance of the extended models has been found high enough to claim a detection. Still, it is becoming clear that hints of new physics (if any) are most likely to be detected in the correlation between different observables.

In this spirit, this is the first in a series of papers in which we search for transient reductions in the speed of sound of the adiabatic mode consistent with (effectively) single-field inflation and \emph{uninterrupted} slow-roll. We do this by exploiting a very simple correlation between power spectrum and bispectrum noted in \cite{Achucarro:2012fd}. While more general situations are possible, and have been considered elsewhere \cite{Adshead:2011bw,Adshead:2013zfa}, there is a particularly interesting regime for which the \emph{complete} primordial bispectrum is obtained to leading order in slow-roll \cite{Achucarro:2012fd}. The amplitude and the rate of change of the speed of sound must be large enough to dominate over slow-roll effects while being small enough to allow a perturbative calculation of the effect on the power spectrum and bispectrum.

Our test case consists of a gaussian reduction in the speed of sound occurring within the window of e-folds in which the scales corresponding to the angular scales probed by Planck exit the Hubble sound horizon. The functional form is inspired by soft turns along a multi-field inflationary trajectory with a large hierarchy of masses, a situation that is consistently described by an effective single-field theory \cite{Achucarro:2010jv,Achucarro:2010da,Cespedes:2012hu,Achucarro:2012sm} (see also \cite{Gao:2012uq,Gao:2013ota}). Nevertheless we stress that reductions in the speed of sound are a more general phenomenon within effective field theory (and hence may have diverse physical origins).

Our statistical analysis of the Planck CMB power spectrum reveals several fits with a moderately improved likelihood compared to the best \lcdm fit. For each of those fits we give the associated full primordial bispectrum. The Planck bispectrum data have not yet been released but, due to a lucky coincidence, templates very similar to our predictions have already been tested by Planck \cite{Ade:2013ydc} (inspired by a step in the potential). We find  that the predicted bispectra for some of our fits are reasonably consistent with the best fits of Planck. In addition, some of our best fits lie on a region of the parameter space not yet analyzed by Planck. If confirmed, these correlations would constitute evidence for transient reductions in the speed of sound. It is interesting that rather mild reductions of the order of a few percent may already be observable in the data.

%%%%%%%%%%%%%%%%%%%%%%%%%%%%%%%%%%%%
%%%%%%%%%%%%%%%%%%%%%%%%%%%%%%%%%%%%%%%
%%%%%%%%%%%%%%%%%%%%%%%%%%%%%%%%%%%%%%%%%%

\section{Correlated features in the primordial spectra from a transient reduction in the speed of sound}

The quadratic action of a general single-field theory for the adiabatic curvature perturbation $\mathcal{R}$ is
\begin{equation}
\label{correlated1}
S_2 = m_\text{Pl}^2\int \mathrm{d}^4x\,a^3\eps\left[\dot{\mathcal{R}}^2-\frac{\left(\nabla\mathcal{R}\right)^2}{a^2}\right]
+m_\text{Pl}^2\int \mathrm{d}^4x\,a^3\eps\left(\frac{1}{c_s^2}-1\right)\dot{\mathcal{R}}^2\ .
\end{equation}
where $c_s$ is the sound speed. The mode functions are easily found for the free ($c_s=1$) action in the first line. Using the in-in formalism \cite{Keldysh:1964ud,Weinberg:2005vy}, the change in the power spectrum due to a small transient reduction in the speed of sound, to first order in $u\equiv1-c_s^{-2}$, is found to be \cite{Achucarro:2012fd}
\begin{equation}\label{eq:deltappfourier}
\frac{\Delta \mathcal{P_R}}{\mathcal{P_R}}(k)=k\int_{-\infty}^{0}\mathrm{d}\tau\ u(\tau)\sin{(2k\tau)}\ ,
\end{equation}
where $k\equiv|\bm{k}|$, $\mathcal{P_R}=H^2/(8\pi^2\epsilon m_{\text{Pl}}^2)$ is the featureless power spectrum with $c_s=1$, and $\tau$ is the conformal time. Here we see how changes in the speed of sound, independently of their physical origin, seed features in the power spectrum. However, different inflationary scenarios will give different coefficients for the cubic operators in the action, and therefore will in general be distinguishable at the level of the bispectrum \cite{Cheung:2007st, Achucarro:2012sm}.

This method provides a clear advantage with respect to those in which the mode functions are calculated from the complete equations of motion \cite{Stewart:2001cd,Adshead:2011jq,Hu:2011vr,Park:2012rh,Bartolo:2013exa}, where higher derivatives of $c_s$ appear and extra hierarchies must be usually imposed. We have checked that both methods agree for sudden variations of the speed of sound \cite{companion}. It is however important to note that \eqref{eq:deltappfourier} assumes $c_s=1$ in the far past ($\tau=-\infty$) and at the end of inflation ($\tau=0$).

One can also calculate the bispectrum disregarding slow-roll contributions $\mathcal{O}(\epsilon,\eta)$ with respect to $u$ and $s\equiv\dot{c_s}/Hc_s$, which ensures that the standard slow-roll result \cite{Maldacena:2002vr} for $c_s=1$ is subdominant with respect to this leading contribution, given by (see \cite{Achucarro:2012fd} for details):
\begin{equation}\label{eq:deltaB}
\Delta B_\mathcal{R}(\bm{k}_1,\bm{k}_2,\bm{k}_3) = \left[
c_0^\triangle(\bm{k}_i)\,\frac{\Delta \mathcal{P_R}}{\mathcal{P_R}}\right.\\
\left.
+c_1^\triangle(\bm{k}_i)\,\frac{\mathrm{d}}{\mathrm{d}k}\left(\frac{\Delta \mathcal{P_R}}{\mathcal{P_R}}\right)
+c_2^\triangle(\bm{k}_i)\,\frac{\mathrm{d}^2}{\mathrm{d}k^2}\left(\frac{\Delta \mathcal{P_R}}{\mathcal{P_R}}\right)
\right]\bigg|_{k= \sum \frac{|\bm{k}_i|}{2}}\ .
\end{equation}

%%%%%%%%%%%%%%%%%%%%%%%%%%%%%%%%%%
%%%%%%%%%%%%%%%%%%%%%%%%%%%%%%%%%%%
%%%%%%%%%%%%%%%%%%%%%%%%%%%%%%%%%%%%%

In this work, we choose to parametrize the reduction in the speed of sound as a gaussian in e-folds $N$ as follows:
\begin{equation}
u=1-c_s^{-2}=B\,e^{-\beta(N-N_0)^2}=B\,e^{-\beta\left(\ln\frac{\tau}{\tau_0}\right)^2}\ ,
\label{eq:gaussefolds}
\end{equation}
where $\beta>0$, $B<0$ and $N_0$ (or $\tau_0$) is the instant of maximal reduction. Assuming slow-roll, $\ln\left(-\tau\right)=\left(N_\text{in}-N\right)-\ln\left(a_\text{in}H_0\right)$, where $a_\text{in}=a(N_\text{in})$ and $N_\text{in}$ is the time when the last $\sim$ 60 e-folds of inflation start.

The angular scales probed by Planck ($\ell=2-2500$) correspond to certain scales in momentum space crossing the Hubble horizon during the first $N_\text{CMB}\simeq$ 7 e-folds of the last $\sim$ 60 e-folds of inflation. The range of $N_0$ and the lower bound on $\beta$ are chosen to give a reduction of the speed of sound well contained within this CMB window. The range of $B$ and the upper bound $\beta$ must be such that the perturbative calculations are valid  and the rate of change of the speed of sound is small. We take $|u|,|s|\ll1$. Altogether, the allowed region of our parameter space is taken to be \cite{companion}:
\begin{subequations}
\label{eq:bounds}
\begin{gather}
\mathcal{O}(\epsilon,\eta)    \ll|B|\ll         1\ ,\label{eq:bound1}\\
\frac{50}{N_\text{CMB}^2}<\beta\ll         \frac{2e}{B^2}\ ,\label{eq:bound2}\\
\frac{5}{\sqrt{2\beta}}    <N_0-N_\text{in}< N_\text{CMB}-\frac{5}{\sqrt{2\beta}}\ .\label{eq:bound3}
\end{gather}
\end{subequations}
This is a very conservative choice. First, \eqref{eq:bound3} and the lower bound in \eqref{eq:bound2} are more restrictive than the condition that the feature be observable. For example, we expect observable effects when the reduction occurs before the CMB window, since it would effectively modify the initial conditions of the modes subsequently leaving the sound horizon. We are also trying to avoid very broad features that could be degenerate with cosmological parameters as the spectral index $n_s$ and the optical depth $\tau_{\text{reio}}$, as well as highly oscillating features (for large values of $|\tau_0|$) that make computational control difficult.

Secondly, this range is well within the region of the parameter space where the cubic lagrangian is much smaller than the quadratic lagrangian, and hence is perturbatively under control. An extension to the full perturbative region is currently under investigation \cite{follow-up}.

%%%%%%%%%%%%%%%%%%%%%%%%%%%%%%%%
%%%%%%%%%%%%%%%%%%%%%%%%%%%%%%%%%%
%%%%%%%%%%%%%%%%%%%%%%%%%%%%%%%%%%%%

\section{Methodology of the search}
We consider features from a transient reduction in the speed of sound described by the ansatz \eqref{eq:gaussefolds}. For its three parameters, we take uniform priors on $B$, $\ln\beta$ and $\ln(-\tau_0)$. Their ranges are given by eqs.\ \eqref{eq:bounds} and a stronger restriction than \eqref{eq:bound3}
\begin{equation}
4.4 < \ln(-\tau_0) < 6\ ,\label{eq:2bound3}
\end{equation}
which is motivated by a search for bispectrum features by the Planck collaboration \cite[sec.\ 7.3.3]{Ade:2013ydc}. The model-dependent bound $|B|\gg\mathcal{O}(\epsilon,\eta)$ is ignored a priori.

The primordial power spectrum feature at eq.\ \eqref{eq:deltappfourier} is computed using a Fast Fourier Transform, and added to the primordial spectrum of the \lcdm Planck baseline model described in ref.\ \cite[sec.\ 2]{Ade:2013zuv}. The resulting CMB power spectrum, calculated using the \class Boltzmann code \cite{Lesgourgues:2011re,Blas:2011rf}, is fitted to the Planck CMB temperature data \cite{Planck:2013kta} and the WMAP CMB low-$\ell$ polarization data \cite{Bennett:2012zja}, using \montepython \cite{Audren:2012wb} as a Markov chain Monte Carlo (MCMC) sampler. We varied all cosmological, nuisance and feature parameters. For those last ones, the likelihood probability distribution is found to be multi-modal. Though multi-modal distributions are more efficiently sampled using other methods (e.g.\ \codename{MultiNest} \cite{Feroz:2007kg, Feroz:2008xx}), we were able to perform the search using only MCMC's (see Appendix A for details on the methodology of the search).

%%%%%%%%%%%%%%%%%%%%%%%%%%%%%%%%%%%%%%%%
%%%%%%%%%%%%%%%%%%%%%%%%%%%%%%%%%%%%%%%%%
%%%%%%%%%%%%%%%%%%%%%%%%%%%%%%%%
%%%%%%%%%%%%%%%%%%%%%%%%%%%%%%%%%

\section{Summary of  results}

\begin{figure}
\centering
\includegraphics[width=0.8\columnwidth]{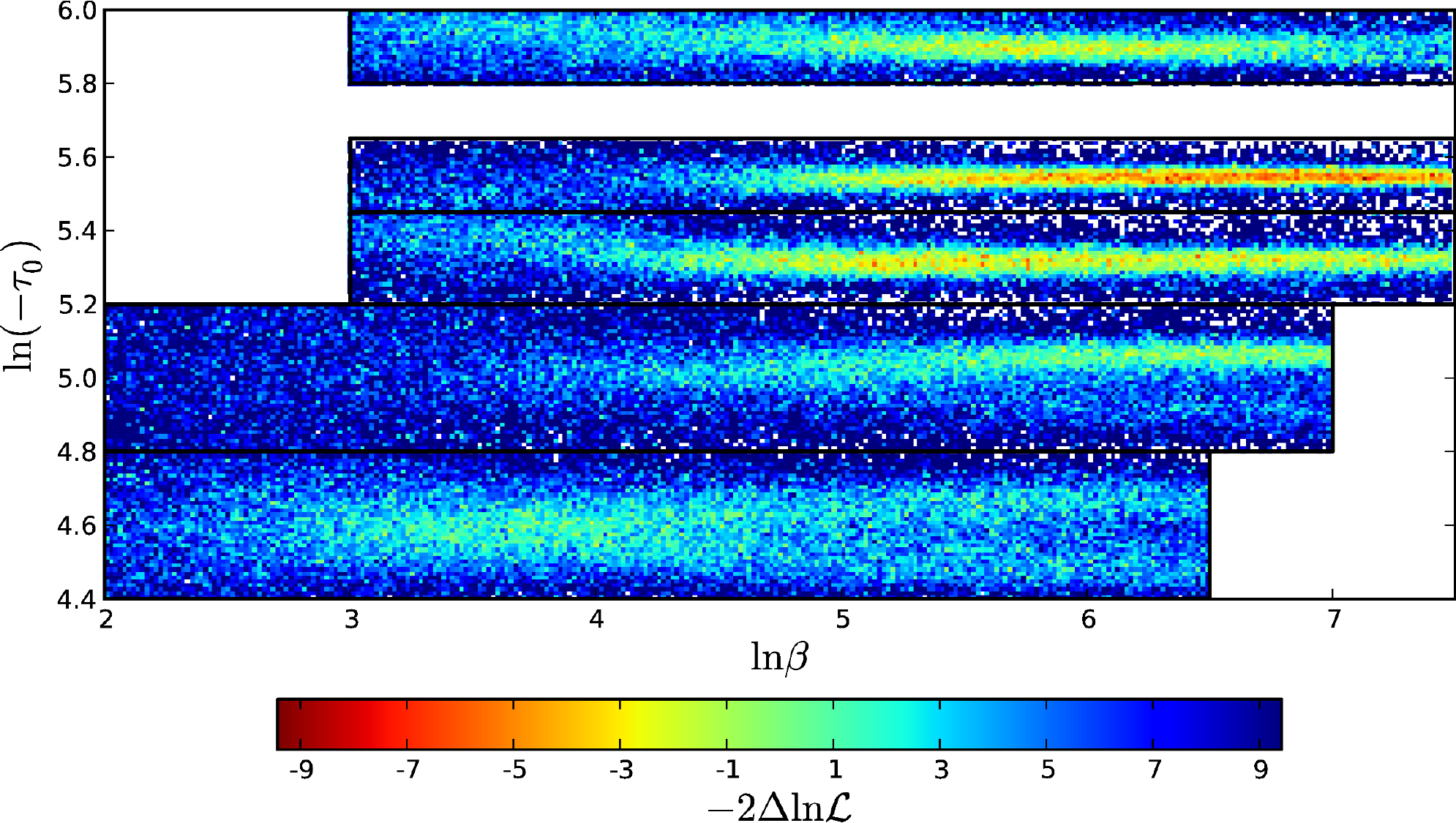}
\caption{\label{fig:modes}
Profile of $\Delta\chi^2_\text{eff}=-2\Delta\ln\mathcal{L}$ for the features in the CMB power spectrum in the $\left(\ln\beta,\,\ln(-\tau_0)\right)$ plane.}
\end{figure}

The result of our search, having discarded small signals with\footnote{Hereafter, $\chi^2$ refers to the \emph{effective} quantity defined as $\chi^2_\text{eff} = -2\ln\mathcal{L}$, see \cite[p.\ 10]{Verde:2009tu}; in turn, $\Delta$ stands for the difference with the corresponding best fit value of Planck baseline model, using the same likelihood.} $\Delta\chi^2 > -2$ over \lcdm, is a series of five well-isolated bands of almost constant $\ln(-\tau_0)$, with variable significance, see table \ref{tab:paramranges} and figure \ref{fig:modes}.

The amplitude $B$ of the fits is rather small, $\mathcal{O}(10^{-2})$, and therefore comparable with neglected slow-roll terms. This means the bispectrum is dominated by terms of order $s  = \dot{c_s}/(Hc_s)$. The maximum values of $s$ at the best fits for the modes \mode{A} to \mode{E} in table \ref{tab:paramranges} are respectively $0.33$, $0.42$, $0.40$, $0.48$, $0.05$. Notice that the value of $s$ for \mode{E} is also comparable to neglected terms, so the prediction for the bispectrum based on eq.\ \eqref{eq:deltaB} cannot be trusted in this case. We therefore disregard this mode in the comparison with the bispectrum.

\begin{table}
\newcommand{\btf}[1]{\ensuremath{(#1)}}
\newcommand{\btfint}[4]{$\btf{#1}\,#2\,\substack{+#3\\-#4}$}
\newcommand{\btfintsimple}[3]{$\btf{#1}\,#2\,\pm #3$}
\newcommand{\colspacing}{~~}
\centering
\begin{tabular}{c@{\colspacing}c@{\colspacing}c@{\colspacing}c@{\colspacing}c}
\# & $-B\times 10^{2}$       & $\ln\beta$ & $\ln(-\tau_0)$ & $\Delta\chi^2$\\
\hline\hline
\mode{A} & \btfint{4.5}{3.7}{1.6}{3.0}
  & \btfint{5.7}{5.7}{0.9}{1.0}
  & \btfint{5.895}{5.910}{0.027}{0.035}
  & $-4.3$\\[1mm]
\hline
\mode{B} & \btfintsimple{4.2}{4.3}{2.0}
  & \btfint{6.3}{6.3}{1.2}{0.4}
  & \btfint{5.547}{5.550}{0.016}{0.015}
  & $-8.3$\\[1mm]
\hline
\mode{C} & \btfint{3.6}{3.1}{1.6}{1.9}
  & \btfint{6.5}{5.6}{1.9}{0.7}
  & \btfint{5.331}{5.327}{0.026}{0.034}
  & $-6.2$\\[1mm]
\hline
\mode{D} & \btf{4.4}
  & \btf{6.5}
  & \btf{5.06}
  & $-3.3$\\[1mm]
\hline
\mode{E}$^*$ & \btf{1.5}
  & \btf{4.0}
  & \btf{4.61}
  & $-2.2$
\end{tabular}
\caption{\label{tab:paramranges}
CMB power spectrum best fits (in parentheses), $68\%$ c.l.\ intervals and effective $\Delta\chi^2$ at the best fit value for each of the different modes. The prediction for the bispectrum for \mode{E} is not reliable (see text).}
\end{table}

For the modes \mode{A}, \mode{B} and \mode{C} the table shows the $68\%$ c.l.\ ranges. For bands \mode{B} and \mode{C} we were unable to put an upper bound on $\ln\beta$  due to a degeneracy between that parameter and the amplitude $|B|$. For those two modes, the upper bound on $\ln\beta$ is set by the prior $s<1$ in eq.\ \eqref{eq:bound2}, which is saturated at $\ln\beta \simeq 7.5$. The best fit for \mode{B} lies at $s \simeq 1$, so we present in table \ref{tab:paramranges} the second best (see Appendix B for the predictions of mode \mode{B} and an illustration of the $(B, \ln\beta)$ degeneracy). The high-$\ell$ CMB polarization data of the upcoming 2.5 years data release of Planck should put an upper bound on $\ln\beta$, as well as confirm that we are not fitting noise.

The lower bands \mode{D} (and \mode{E}) are less significant and their likelihoods much less gaussian, so we only show their best fits (for parameter constraints see \cite{companion}). Despite their low significance, they are worthy of mention because they fall in the region overlapping with Planck's search for features in the bispectrum (see below).

The best fits and $68\%$ c.l.\ ranges \cite{Ade:2013zuv} of the six \lcdm parameters are quite accurately reproduced. We find two mild degeneracies ($|r|\lesssim 0.15$) of $\ln(-\tau_0)$ with $\omega_\text{CDM}$ and $H_0$ \cite{companion}. Best fits and confidence intervals are also preserved for the nuisance parameters. The study of a possible degeneracy with the lensing amplitude is left for future work.

A gain of $|\Delta\chi^2|\lesssim 10$ is common in similar searches (see Appendix C for a comparison with other searches for features in the CMB power spectrum), which suggests that CMB power spectrum data alone cannot justify the introduction of these features. Nevertheless, the aim of this paper is to show that low-significance fits can still predict correlated features in the bispectrum which are possibly observable with the current data. Model selection should be done taking into account both observables (or naturally, any other combination).

%%%%%%%%%%%%%%%%%%%%%%%%%%%%%%%%
%%%%%%%%%%%%%%%%%%%%%%%%%%%%%%%%
%%%%%%%%%%%%%%%%%%%%%%%%%%%%%%%%

\section{Comparison with the search for features in Planck's bispectrum}

A search for linearly oscillatory features was performed on Planck's bispectrum (cf. \cite[sec.\ 7.3.3]{Ade:2013ydc}), using as a template \cite{Chen:2006xjb}
\begin{equation}\label{eq:Btestmodel}
B(k_1, k_2, k_3) = \frac{6A^2f_\text{NL}^\text{feat}}{(k_1k_2k_3)^2}
                  \sin\left(2\pi\frac{\sum_{i=1}^3 k_i}{3k_c}+\phi\right)
\ ,
\end{equation}
where $A = A_sk_*^{1-n_s}$, $A_s$ and $n_s$ being the amplitude and spectral index of the primordial power spectrum, and $k_*=\unit[0.05]{Mpc^{-1}}$ a pivot scale. They sampled the amplitude $f_\text{NL}^\text{feat}$ over a coarse grid of wavelengths $k_c$ and phases $\phi$. 

Our features also present a linearly oscillatory pattern, which comes from the Fourier transform in \eqref{eq:deltappfourier}. These oscillations enter the bispectrum approximately as $\exp(i\sum_i k_i\tau_0)$, cf.\ eq.\ \eqref{eq:deltaB}, which compares to Planck's search as $\tau_0 \approx 2\pi/(3k_c)$. Thus, Planck's search falls inside $\ln(-\tau_0)\in[4.43,5.34]$, while ours spans up to $\ln(-\tau_0) = 6\,\left(k_c = \unit[0.00519]{Mpc^{-1}}\right)$. The overlap includes our modes \mode{C} and \mode{D} (and also the discarded \mode{E}).

The search in \cite{Ade:2013ydc} is later supplemented with a gaussian envelope centered at scales corresponding to the first acoustic peak, which dampens the signal in subsequent peaks for decreasing values of a falloff\footnote{James Fergusson, private communication.} $\Delta k$. The envelope generally improves the significance, except for the $2\sigma$ signal at $k_c = 0.01375\,,\unit[0.01500]{Mpc^{-1}}$. This suggests that this band's significance comes mostly from the second and third peaks (the signal from the fourth on would be most likely damped out).

\begin{figure}
\centering
%\unit[0.01375]{Mpc^{-1}}
\subfigure[Comparison of Planck's CMB power spectrum (blue) and the corresponding best fit of the mode \mode{D} (red).]{\includegraphics[width=0.70\columnwidth]{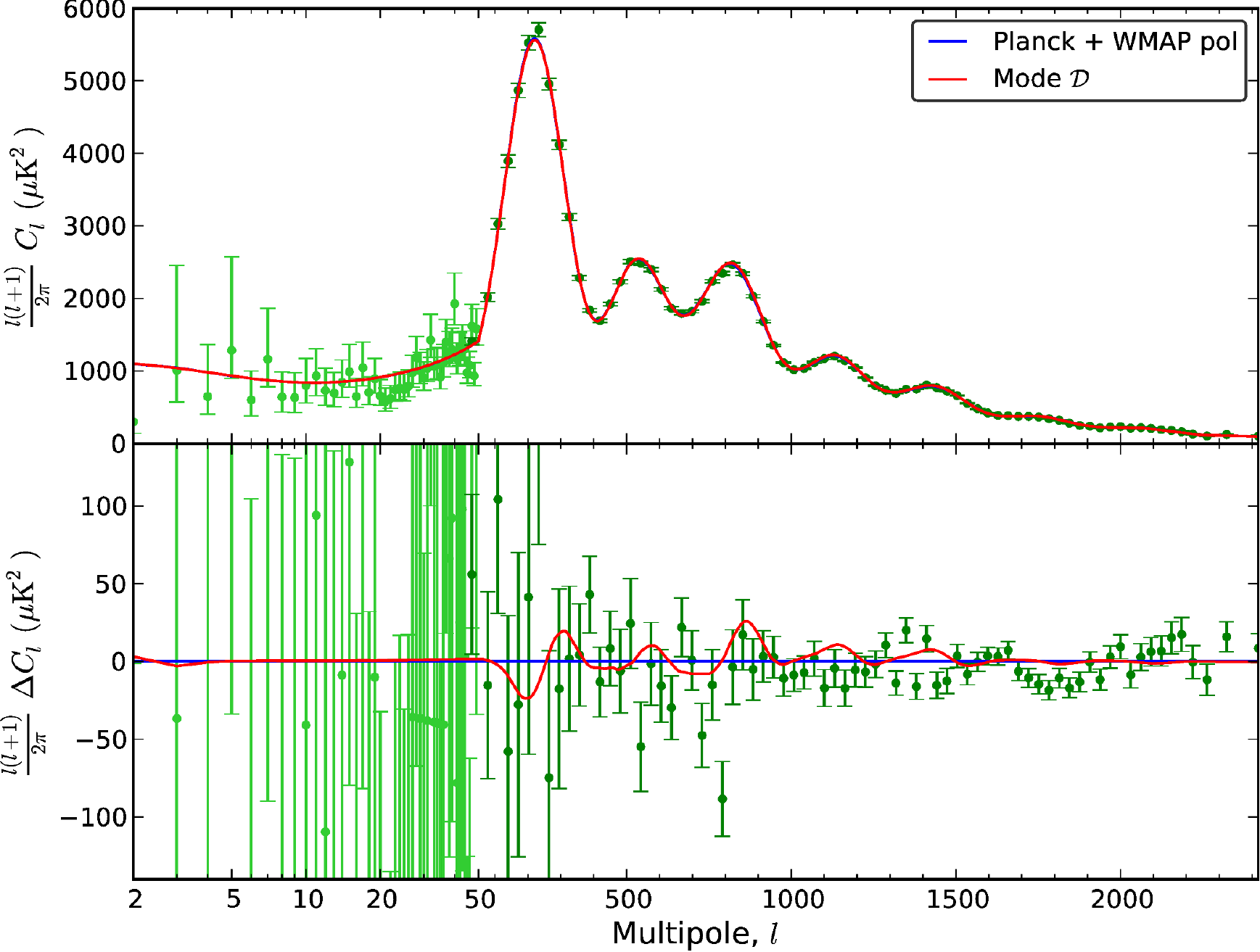}}
\subfigure[Comparison along the equilateral direction of Planck's $2.3\sigma$ \emph{primordial} bispectrum fit with $k_c = 0.01375\,\mathrm{Mpc}^{-1}$ (dashed), and the expected signal in the \emph{primordial} bispectrum for the best fit of \mode{D} (solid). Both bispectra are normalized by $f(k_1,k_2,k_3) = (10/3)\,\left((2\pi)^2 A_sk_*^{1-n_s}\right)^{-2} \prod_ik_i^3\,/\,\sum_ik_i^3$. The gray stripes show the approximate scales corresponding to the first four acoustic peaks in the CMB power spectrum. Although our signal extends beyond those scales (see zoom-out at the lower-left corner), from the third peak on, the primordial signal is highly suppressed by diffusion damping when transferred to the CMB.]{\includegraphics[width=0.70\columnwidth]{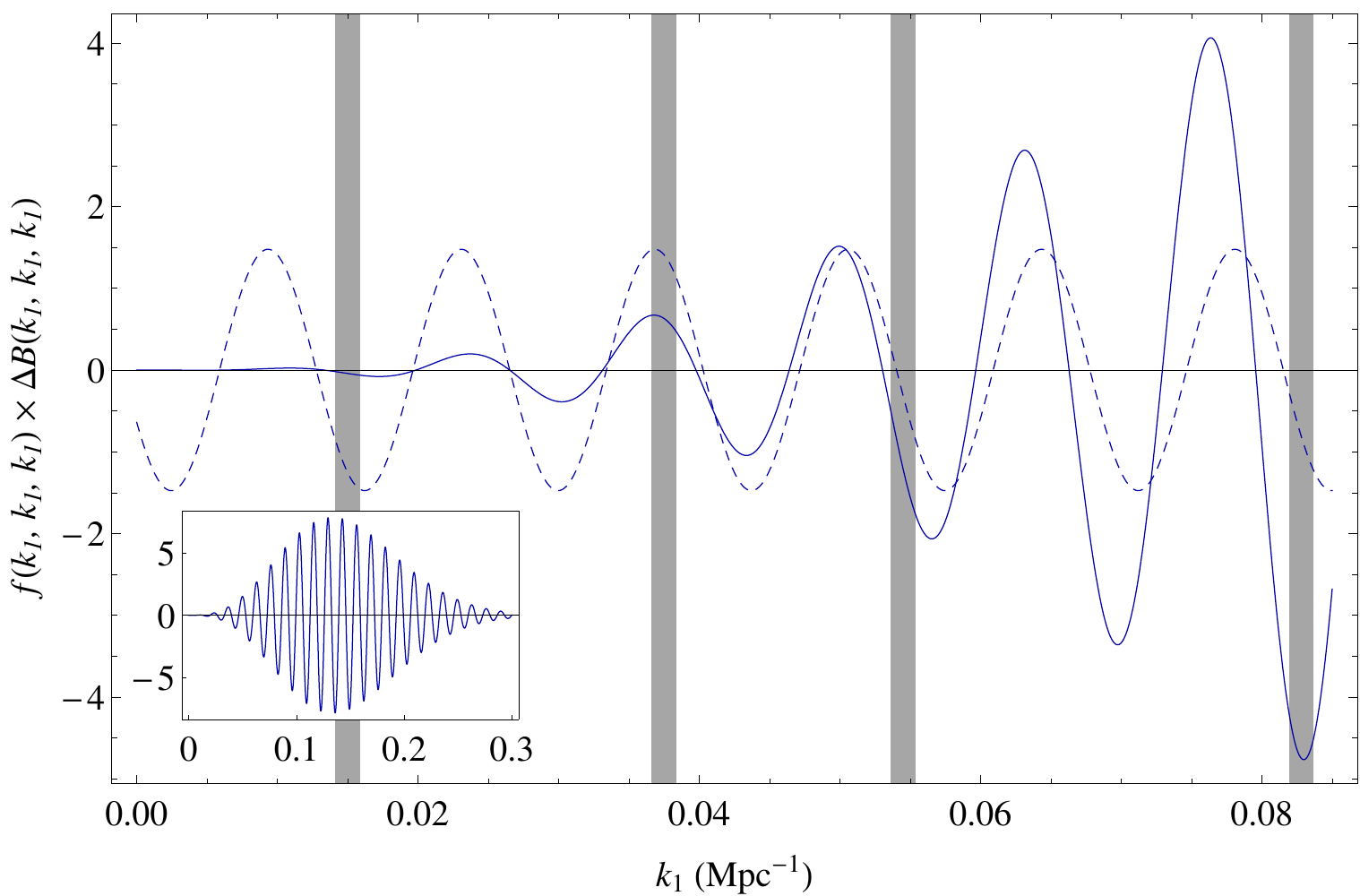}}
\caption{\label{fig:bicomp}
Features corresponding to the best fit of the mode \mode{D} (see table \ref{tab:paramranges}), for which the comparison with Planck analysis for the bispectrum is possible.}
\end{figure}

In comparison, our best fits to the power spectrum predict bispectrum features which are mild at the first peak and more intense from the second peak onwards. The higher the value of $\ln \beta$, the smaller the scale at which the feature peaks. 
In the range of $\ln(-\tau_0)$ probed here, we were not able to reproduce the improvement Planck appears to see for features at the first peak. 
On the other hand, we find good matching around the second and third peak scales between the best fit of \mode{D} with $k_c = \unit[0.01327]{Mpc^{-1}}$ and the $2.3\sigma$ signal of Planck at $k_c = \unit[0.01375]{Mpc^{-1}}$ with $f_\text{NL}^\text{feat} = 345$ and $\phi = \pi/2$ (see fig.\ \ref{fig:bicomp}). A milder matching also occurs at the same scales between the best fit of \mode{C} with $k_{c} = \unit[0.01014]{Mpc^{-1}}$ and Planck's $2.6\sigma$ signal with\footnote{Note that we have quoted the fits to the Bispectrum of Planck without applying the \emph{look-elsewhere effect}. This effect will be properly taken into account when a full study of the Bayesian evidence is performed in a future work.} $k_c = \unit[0.01125]{Mpc^{-1}}$.

Although this matching is not easy to quantify, it suggests enlarging the search in \cite{Ade:2013ydc} to cover the frequencies corresponding to modes \mode{A} and  \mode{B}, and to test envelopes centered at smaller scales.

%%%%%%%%%%%%%%%%%%%%%%%%%%%%%%%%%%
%%%%%%%%%%%%%%%%%%%%%%%%%%%%%%%%%%%%
%%%%%%%%%%%%%%%%%%%%%%%%%%%%%%%%%%%%%%%

\section{Conclusions and discussion}

We have carried out a statistical search for localized oscillatory features in the CMB power spectrum produced by a transient reduction in the speed of sound. We have found a number of fits and calculated the associated primordial bispectra. Because of the small amplitude at the best fits, the bispectrum prediction closely resembles that of step inflation, tested by the Planck collaboration, since a transient slow-roll violation switches on the same operator in the cubic action. It is then straightforward to compare our prediction with the templates used in that search,
and the agreement is surprisingly good. This is remarkable, considering that these bispectrum features are \emph{predicted from a search in the CMB power spectrum} with a very simple ansatz for $c_s$. 
 
The functional form chosen for the reduction in the speed of sound is inspired by soft turns in a multi-field inflationary trajectory with a large hierarchy of masses, a situation that is consistent with an effectively single-field description with uninterrupted slow-roll. Other functional forms and parameter ranges are under investigation \cite{follow-up}. We stress that our analysis is independent of the physical mechanism behind the reduction.

We emphasize that the CMB power spectrum data alone can hardly justify the introduction of features on top of the \lcdm model; a gain of $|\Delta\chi^2|\lesssim 10$ is not uncommon. However, as we have shown, low-significance fits in the power spectrum can still predict correlated features that may be observable in the CMB bispectrum. Therefore, model selection should take into account both observables simultaneously.

Our results suggest that, by exploiting correlations between different observables, current data might already be sensitive enough to detect transient reductions in the speed of sound as mild as a few percent, opening a new window for the presence of extra degrees of freedom during inflation.

%\vspace{5mm} 
%%%%%%%%%%%%%%%%%%%%%%%%%%%%%%%%%%
%%%%%%%%%%%%%%%%%%%%%%%%%%%%%%%%%%%%
%%%%%%%%%%%%%%%%%%%%%%%%%%%%%%%%%%%%%%%%
 
\subsection{Acknowledgments}

We are grateful to James Fergusson, Bin Hu and Wessel Valkenburg for very useful discussions and suggestions and also to Jinn-Ouk Gong, Daan Meerburg, Gonzalo Palma, Subodh Patil, Bartjan van Tent and Licia Verde. J.T.\ thanks Benjamin Audren and Julien Lesgourgues for their great support using \class and \montepython.
 
This work was partially supported by the Netherlands Foundation for Fundamental Research on Matter F.O.M., the DFG Graduiertenkolleg ``Particle Physics at the Energy Frontier of New Phenomena'', a Leiden Huygens Fellowship and the Netherlands Organization for Scientific Research (NWO/OCW) under the Gravitation Program.

%%%%%%%%%%%%%%%%%%%%%%%%%%%%%%%%%%%%%%%%%
%%%%%%%%%%%%%%%%%%%%%%%%%%%%%%%%%%%%%%%%%
%%%%%%%%%%%%%%%%%%%%%%%%%%%%%%%%%%%%%%%%%

\appendix

\section{Appendix A: Details on the methodology of the search}

The primordial power spectrum features caused by a transient reduction in the speed of sound, eq.\ \eqref{eq:deltappfourier}, are added to the primordial spectrum of the \lcdm Planck baseline model described in ref.\ \cite[sec.\ 2]{Ade:2013zuv}, parametrized by the densities of baryonic and cold dark matter, the current expansion rate, the optical depth due to reionization and the amplitude and spectral index of the spectrum of primordial perturbations. The resulting CMB power spectrum is fitted to the Planck temperature data \cite{Planck:2013kta} and the WMAP low-$\ell$ polarization data \cite{Bennett:2012zja}.

We found the likelihood (and hence the posterior) probability distribution to be multi-modal for the parameters describing the feature. Although multi-modal distributions are sampled more efficiently with methods such as \emph{multimodal nested sampling} \cite{Feroz:2007kg, Feroz:2008xx}, we were able to localize the different modes and split the parameter space into multiple uni-modal distributions using only Markov chain Monte Carlo (MCMC) sampling. We achieved so making use of the \emph{profile likelihood} -- the profile likelihood with respect to a subset $\{\alpha\}$ of the parameters $\{\theta\}$ is $\mathcal{L}(\alpha) =\text{max}_{\{\theta\}-\{\alpha\}}\mathcal{L}(\theta)\,$. We inspected the profile likelihood in the plane $\left(\ln\beta,\,\ln(-\tau_0)\right)$ resulting from long-tailed MCMC's over the feature parameters; it revealed the position and rough size of the different modes, and we used that information to crop uni-modal regions. Finally, the uni-modal regions were sampled separately varying both the feature and the Planck baseline model parameters (and the likelihood's nuisance parameters), in order to obtain definitive posterior probability distributions for the different modes.

%%%%%%%%%%%%%%%%%%%%%%%%%%%%%%%%%%%%%%%%%
%%%%%%%%%%%%%%%%%%%%%%%%%%%%%%%%%%%%%%%%%
%%%%%%%%%%%%%%%%%%%%%%%%%%%%%%%%%%%%%%%%%

\section{Appendix B: The most significant mode in the power spectrum}

\begin{figure}[ht!]
\centering
\includegraphics[width=0.8\columnwidth]{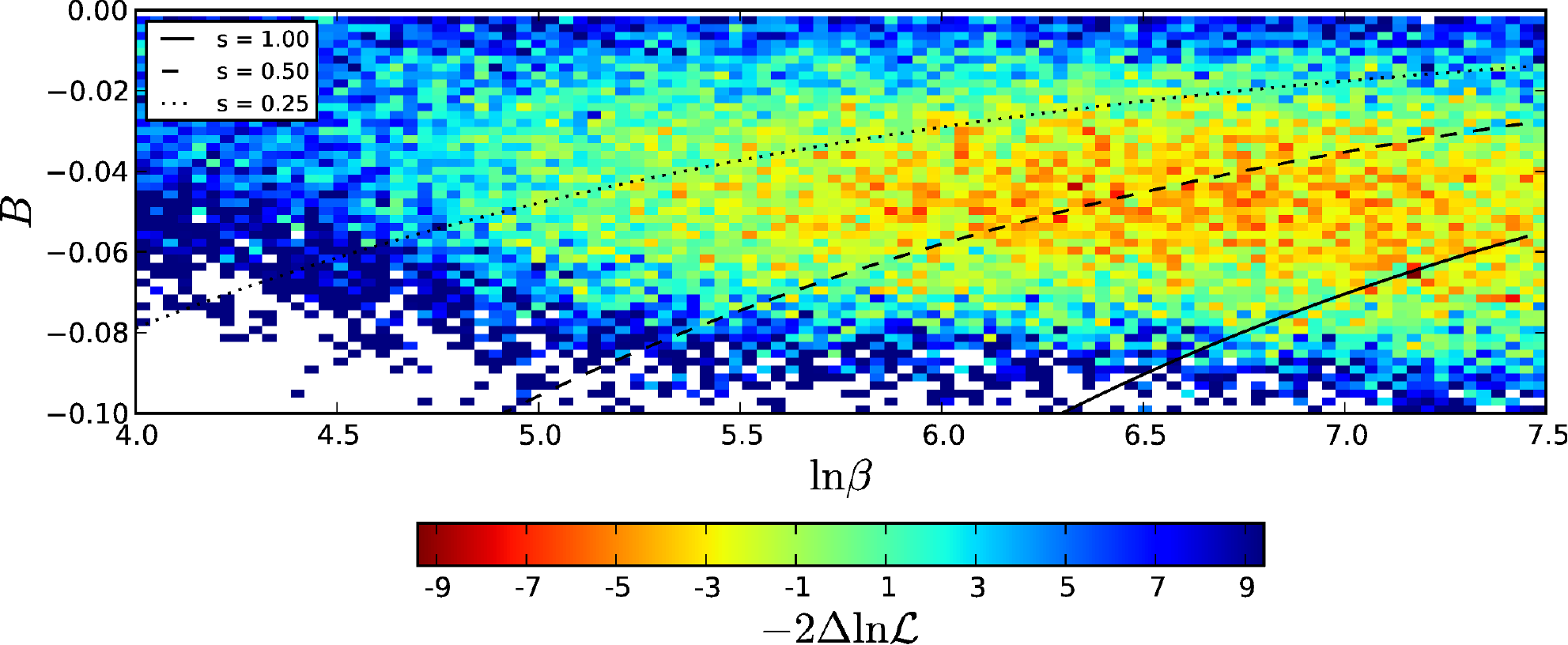}
\caption{\label{fig:Bdeg}
Profile of $\Delta\chi^2_\text{eff}=-2\Delta\ln\mathcal{L}$ for the mode \mode{B} in the $\left(\ln\beta,\,B\right)$ plane, showing the degeneracy between $B$ and $\ln\beta$, and lines of $s=\text{const}$. Notice how the mode saturates the $s<1$ bound.
}
\end{figure}

In this appendix we comment on the characteristics of the mode \mode{B} (see table \ref{tab:paramranges}), which has the highest significance within the region of parameter space considered. As stated in the main text, within this mode (and also in modes \mode{C} and \mode{D}) we find a positive correlation between $\ln\beta$ and $|B|$: along the direction of simultaneous increase of $\ln\beta$ and $|B|$, the feature in the primordial power spectrum broadens towards smaller scales, while the amplitude of the tail on the larger scales remains almost constant. Since the signal at smaller scales will be suppressed in the CMB by diffusion damping, no significance is gained along the degeneracy direction, and this results in a plateau for $\Delta\chi^2$. Along this plateau, the prior limit $s<1$ in eq.\ \eqref{eq:bound2} gets saturated at $\ln\beta  \simeq 7.5$ (see figure \ref{fig:Bdeg}), and hence the prior sets the upper bound for $\ln\beta$. Since the damped temperature signal at small scales ``translates'' into polarization via Thomson scattering, the addition of the high-$\ell$ CMB polarization data of Planck should be able to set an upper bound on $\ln\beta$, as well as to confirm that the enhancement in the likelihood comes not from fitting the noise.

\begin{figure}[ht!]
\centering
\subfigure[CMB temperature power spectrum]{\includegraphics[width=0.475\columnwidth]{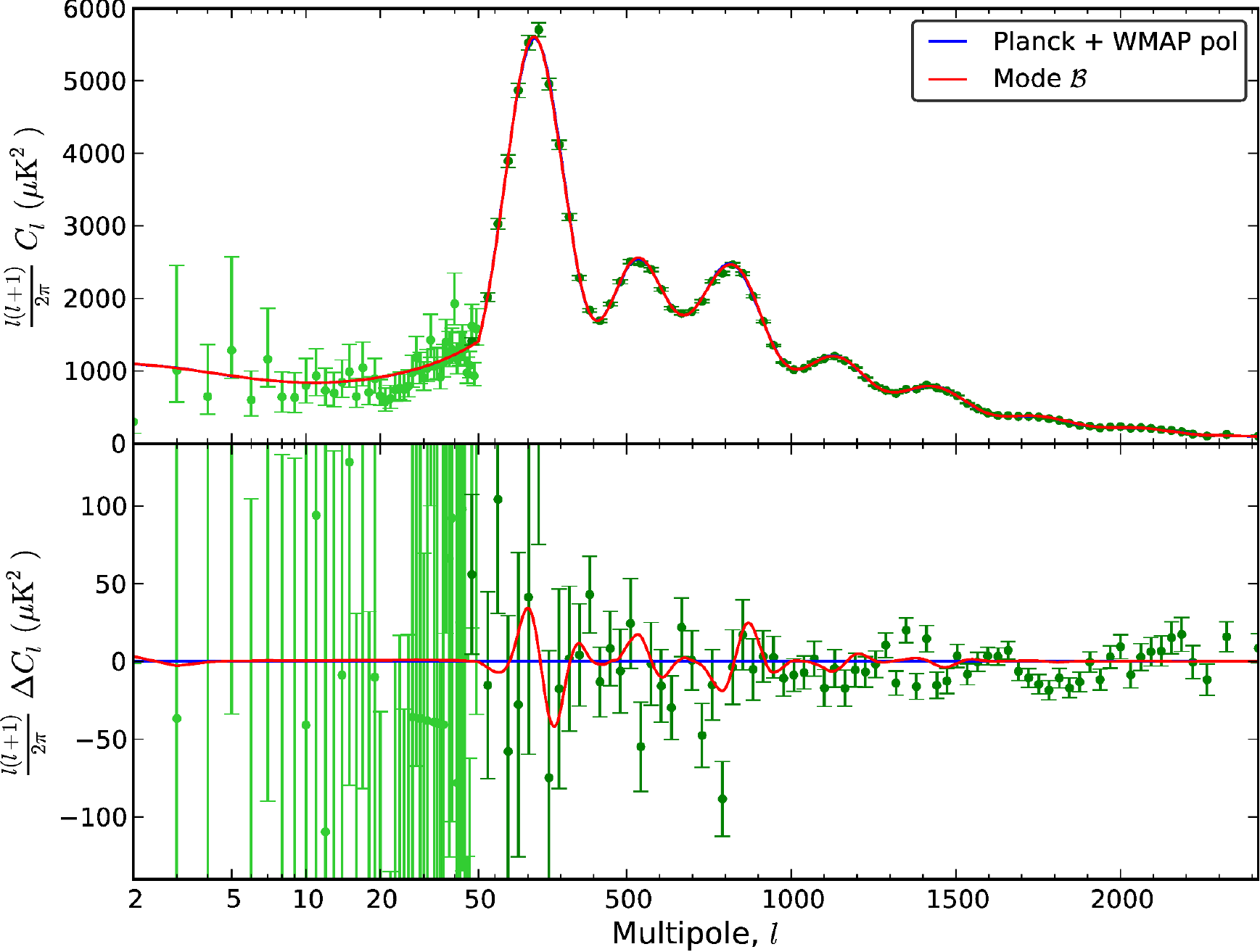}}
\subfigure[CMB TE polarization power spectrum]{\includegraphics[width=0.475\columnwidth]{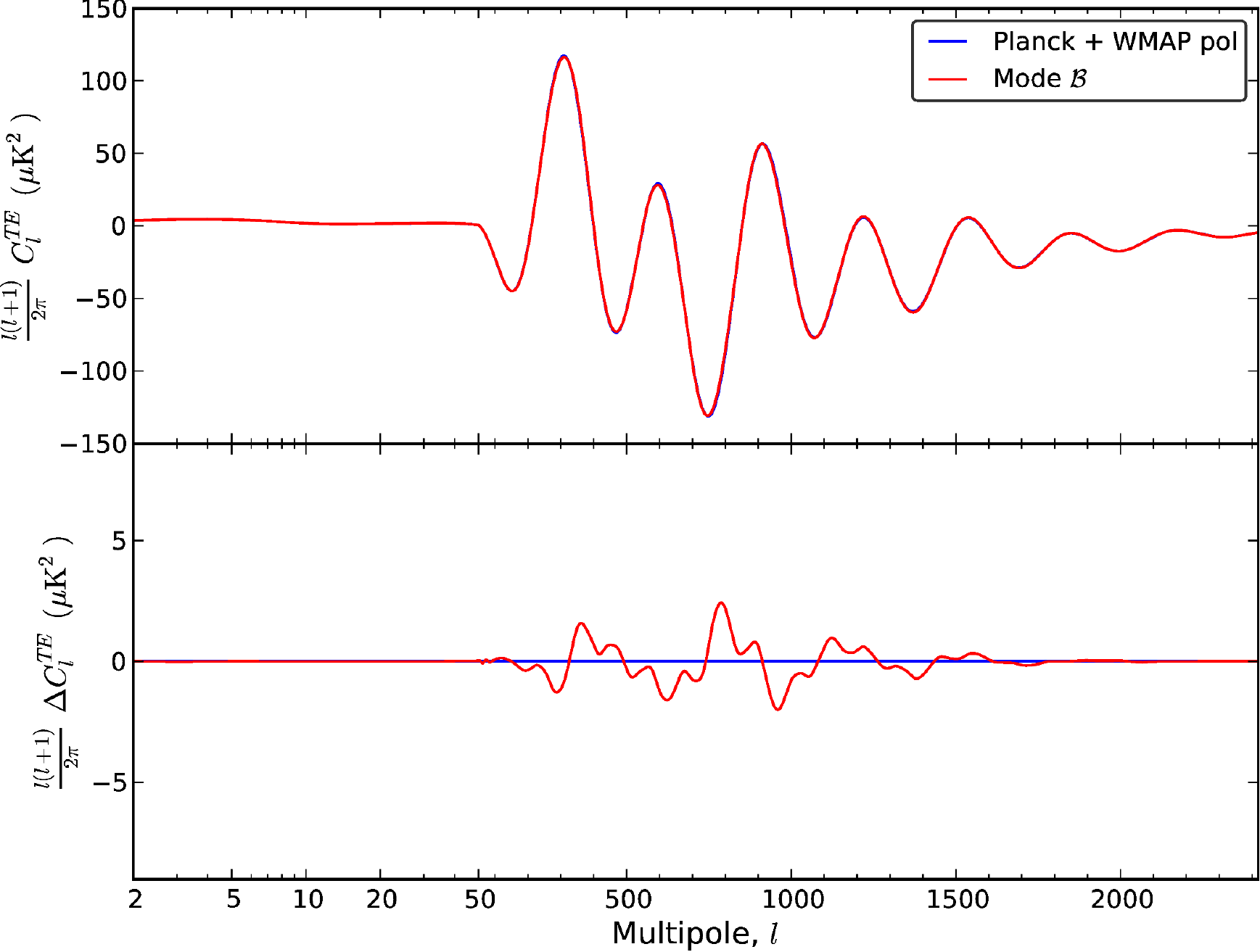}}
\subfigure[CMB EE polarization power spectrum]{\includegraphics[width=0.475\columnwidth]{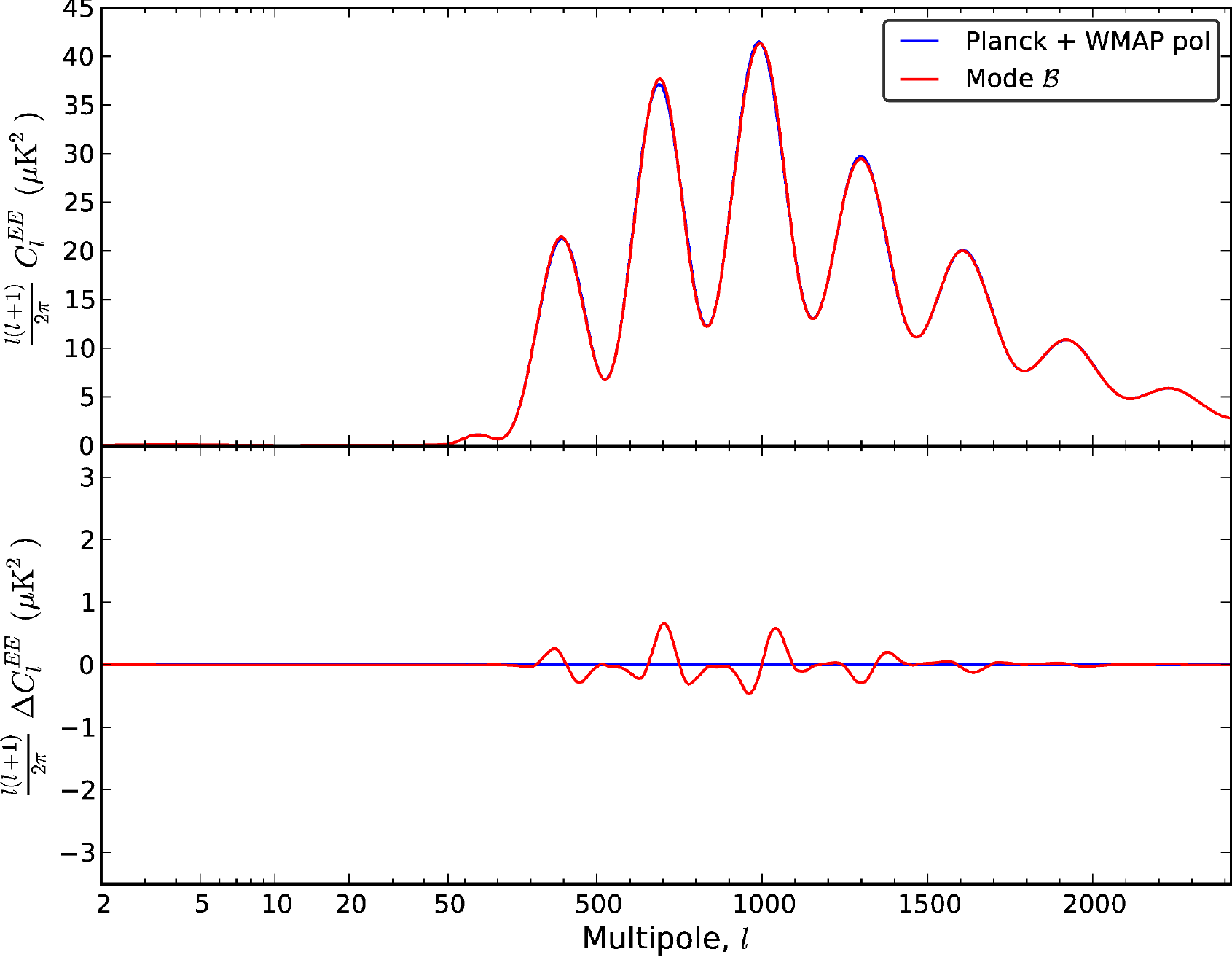}}
\caption{\label{fig:Bclcomp}
Comparison between the CMB temperature and polarization power spectra of Planck (blue) and the corresponding one of the second-best fit of mode \mode{B} (red), see table \ref{tab:paramranges}.
}
\end{figure}

We consider the second-best fit ($\Delta\chi^2 = -8.3$), since the best fit ($\Delta\chi^2 = -9.2$) saturates the prior limit $s<1$ in eq.\ \eqref{eq:bound2}. For the former, we show a comparison with Planck's CMB temperature and polarization power spectra in figure \ref{fig:Bclcomp}.

\begin{figure}[ht!]
\centering
\subfigure[Full 3D primordial bispectrum]{\includegraphics[width=0.475\columnwidth]{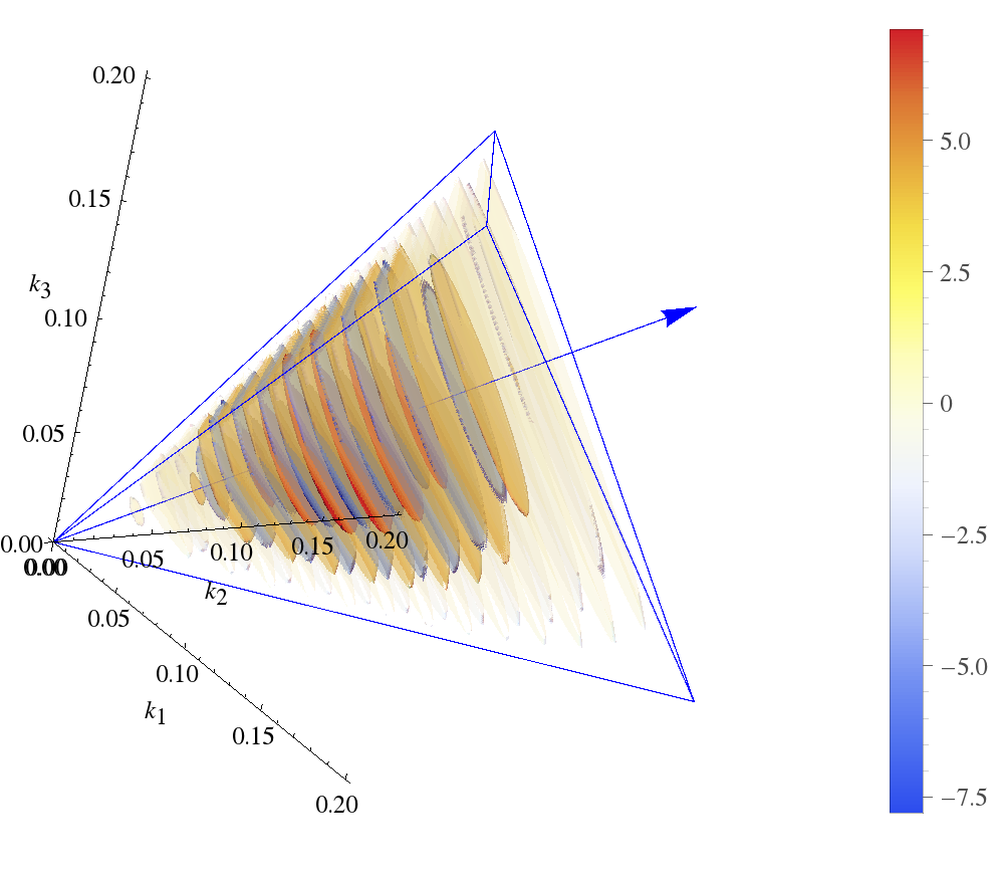}}
\subfigure[Equilateral limit. The gray stripes show the approximate scales of the first four acoustic peaks in the CMB power spectrum, and a zoom-out is shown at the lower-left corner. Most of the signal at high $k$, i.e. small scales, would be suppressed by diffusion damping when transferred to the CMB.]{\includegraphics[width=0.475\columnwidth]{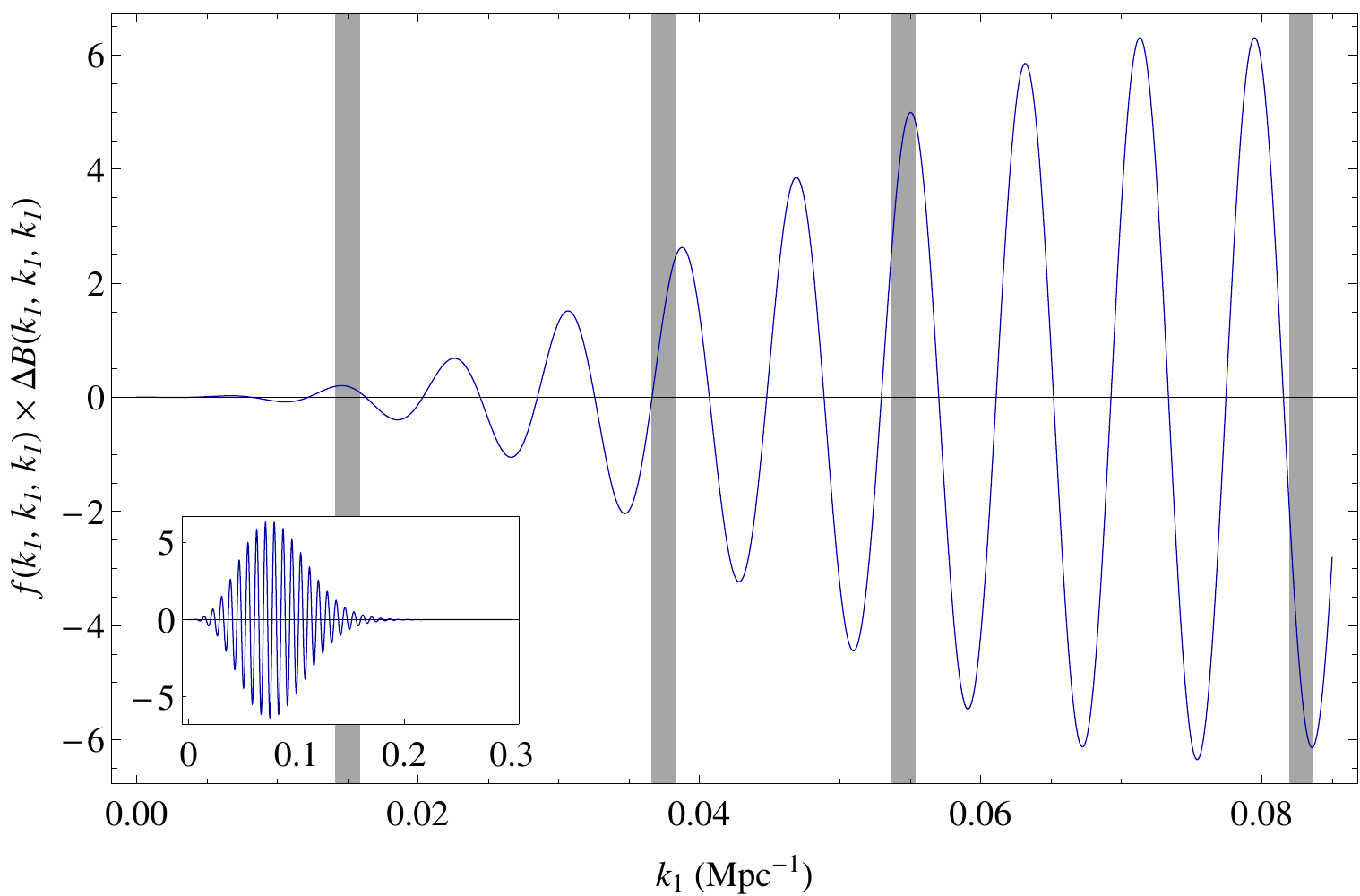}}
\caption{\label{fig:Bbicomp}
Prediction for the \emph{primordial} bispectrum for the second-best fit of mode \mode{B}, normalized by $f(k_1,k_2,k_3) = (10/3)\,[(2\pi)^2 A_sk_*^{1-n_s}]^{-2}
\prod_ik_i^3\,/\,\sum_ik_i^3$.}
\end{figure}

In figure \ref{fig:Bbicomp} we show the prediction for the full primordial bispectrum of the second-best fit.
We expect the signal to be observable in the CMB at scales around the second and third acoustic peaks, since thereafter it will be suppressed by diffusion damping.
In relation to Planck's search in \cite[sec.\ 7.3.3]{Ade:2013ydc}, this feature would be localized at $68\%$ c.l.\ within the interval $k_c\in\left[0.00801,\,0.00826\right]\unit{Mpc^{-1}}$. Thus, testing for it in the current data would require enlarging their search to higher frequencies, i.e.\ smaller values of $k_c$ in eq.\ \eqref{eq:Btestmodel}. Additionally, the significance should be highest when an envelope is placed around the scales corresponding to the second and third peak of the CMB power spectrum.

%%%%%%%%%%%%%%%%%%%%%%%%%%%%%%%
%%%%%%%%%%%%%%%%%%%%%%%%%%%
%%%%%%%%%%%%%%%%%%%%%%%%%%%%%

\section{Appendix C: Comparison with other searches for features in the CMB power spectrum}

Due to the Fourier transform in eq.\ \eqref{eq:deltappfourier}, our features oscillate as $\exp\left(i2k\tau_0\right)$. Thus it is natural to compare to other searches for linearly oscillating features in the Planck CMB power spectrum.

Ref.\ \cite{Meerburg:2013dla} searches for non-localized features with frequencies that compare to ours as $\omega_2=2|\tau_0|$. In the overlapping region, $\omega_2\in[160,\,810]$, they find peaks at roughly $\ln(-\tau_0)\sim\{5.0, 5.1, 5.3, 5.6, 5.7\}\, (|\Delta\chi^2_\text{bf}|\simeq8)$. We find three peaks in this region with similar significance; it could be that the discrepancies come from signals at scales at which our (localized) features are negligible.

Also, the Planck collaboration \cite[sec.\ 8]{Ade:2013uln} searches for features motivated by step-inflation, using the parametrization proposed in \cite{Adshead:2012xz} with a frequency $\eta_f = |\tau_0|$. The profile likelihood in \cite[fig.\ 19, middle]{Ade:2013uln} reveals peaks at $\ln\eta_f\in[4.5,\,4.8]\, (|\Delta\chi^2_\text{bf}|\simeq2)$ and $\ln\eta_f\in[5.3,\,5.7]\, (|\Delta\chi^2_\text{bf}|\simeq8)$, which is consistent with our results.

It is worth noting that in both searches above the overall best fit occurs at $\ln(-\tau_0)\simeq8.2\, (|\Delta\chi^2_\text{bf}|\sim14)$, too high a frequency for the scope of this work.

%\bibliography{features}
%Merlin.mbs v4.21 2009-07-09.
 \newcommand{\noop}[1]{}

\end{document}